\documentclass[journal = aaemcq, manuscript=article, layout=twocolumn]{achemso}

\usepackage{amsmath}
\usepackage{graphicx}
\usepackage{dcolumn}
\usepackage[english]{babel}

\usepackage{color}
\usepackage{float}

\usepackage{amssymb}

\let\oldmaketitle\maketitle
\let\maketitle\relax

\author{Sergei M. Butorin}
\email{sergei.butorin@physics.uu.se}
\affiliation{Condensed Matter Physics of Energy Materials, X-ray Photon Science, Department of Physics and Astronomy, Uppsala University, P.O. Box 516, SE-751 20 Uppsala, Sweden}

\title{Quasi-rigid-band behavior and band gap changes upon isovalent substitution in Cs$_3$Bi$_2$Br$_{9-x}$I$_x$}


\begin{document}

%\date{\today}

%/////////////////////////////////////////////////////////
%-------------------- Abstract ---------------------------
%/////////////////////////////////////////////////////////
\twocolumn[
\begin{@twocolumnfalse}
\oldmaketitle

\begin{abstract}
The recently introduced approach, combining the parameter-free Armiento-K\"{u}mmel generalized gradient approximation exchange functional with the nonseparable gradient approximation Minnesota correlation functional, was used to calculate the electronic structure of the Cs$_3$Bi$_2$Br$_{9-x}$I$_x$ series within density functional theory including the spin-orbit coupling. The changes in the band gap size and its dependence on the $x$ value was investigated. The band gap was found to be of indirect nature and it decreases with increasing I content as long as the system is in the $P\overline{3}m1$ phase. A clear non-linear dependence of the band gap size on $x$ was established which is in qualitative and quantitative agreement with reported experimental data. The quasi-rigid band behavior of the states in the valence and conduction bands of the $P\overline{3}m1$ phase is discussed since no significant changes in the shape of the total density of unoccupied states were observed upon the isovalent substitution.

\end{abstract}

\end{@twocolumnfalse}
]

%\begin{tocentry}
%\includegraphics[height=4.5cm]{TOC.pdf}
%\end{tocentry}

%/////////////////////////////////////////////////////////
%------------------- Introduction ------------------------
%/////////////////////////////////////////////////////////
%\hfill \break
\section{Introduction}

Hybrid metal halide perovskites (HaPs) are considered to be advanced materials for solar cells \cite{Kojima}. The Pb-based HaPs show the most attractive optoelectronic properties in this respect and most often used in the corresponding solar-cell applications. However, the Pb-based HaPs reveal insufficient stability (prone to easy oxidation in air) and contain a toxic element, thus stimulating a search for alternative options. Among those, Bi-based HaPs draw some attention as more stable and environmental-friendly materials. At the same time, all-inorganic halide perovskites are anticipated to show higher stability than the organic-inorganic hybrid ones. In particular, lead-free vacancy-ordered layered perovskites Cs$_3$Bi$_2$I$_9$ and Cs$_3$Bi$_2$Br$_{9}$ are investigated \cite{Machulin,Park,Lehner,Pazoki,Johansson,Huang,Ghosh,Phuyal,Nila,Khazaee,Sun,Gu,Yu,Rieger,Li,Bass,Jin,Biswas} as a suitable replacement. As in case of Pb-based HaPs, the control over the band gap size in the lead-free perovskites is important. This can be achieved by isovalent substitutions of halides, such as e.g. in Cs$_3$Bi$_2$Br$_{9-x}$I$_x$.

\begin{figure*}
\includegraphics[width=\textwidth]{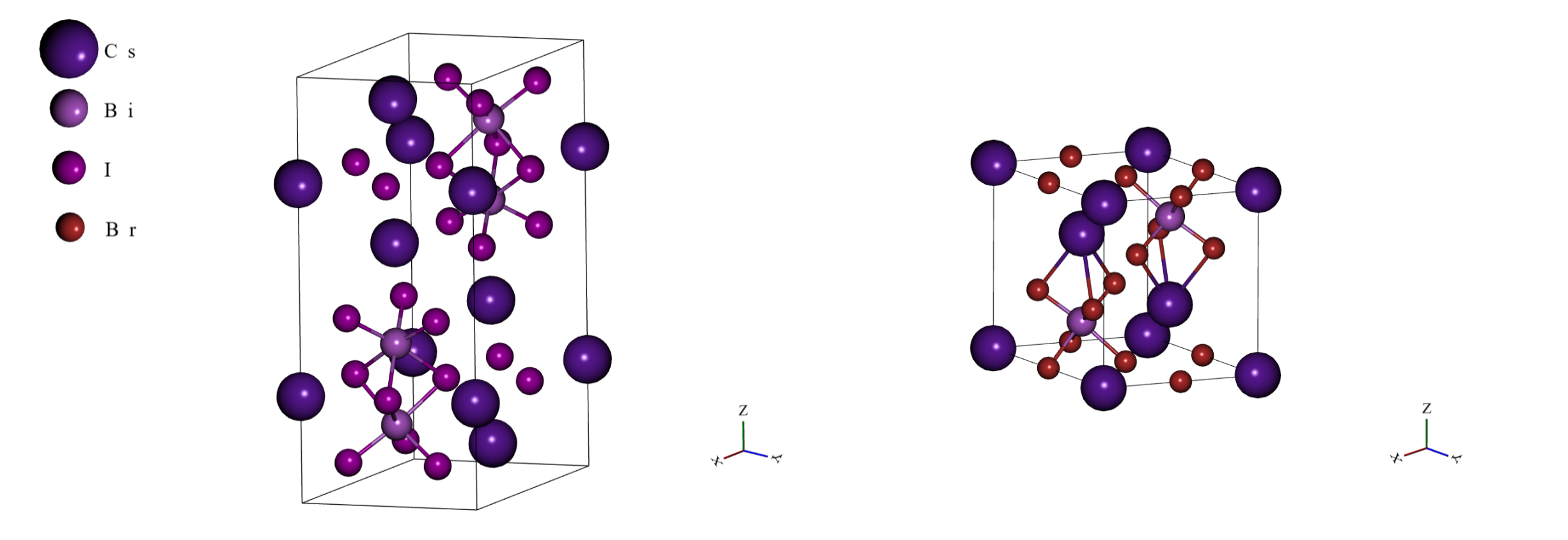}
\caption{Crystal structures of Cs$_3$Bi$_2$I$_9$ ($P6_3/mmc$ type, left side) and Cs$_3$Bi$_2$Br$_9$ ($P\overline{3}m1$ type, right side). \label{Structure}}
\end{figure*}

Cs$_3$Bi$_2$I$_9$ and Cs$_3$Bi$_2$Br$_{9}$ (Fig.~\ref{Structure}) are reported to have the minimum band gap as indirect. While the difference between the sizes of the indirect and direct band gaps in Cs$_3$Bi$_2$Br$_{9}$ were found to be very small (2.57 versus 2.64 eV, respectively, in Ref. \cite{Jin}), such a difference measured to be larger in Cs$_3$Bi$_2$I$_9$ \cite{Ghosh,Gu,Yu}. However, at the same time, there is a large variation in reported values of the indirect band gap in Cs$_3$Bi$_2$I$_9$, ranging from 1.81 to 2.86 eV \cite{Machulin,Park,Lehner,Pazoki,Johansson,Huang,Ghosh,Phuyal,Nila,Khazaee,Sun,Gu,Yu,Li}. This may be connected to the fact that measurements were performed for samples in various forms, such as films, powder, microcrystals and single crystals.

Upon Br substitution by I in Cs$_3$Bi$_2$Br$_{9-x}$I$_x$, the band gap decreases up to $x$=0.6 \cite{Yu,Li,Hodgkins} while the Cs$_3$Bi$_2$Br$_{9-x}$I$_x$ system stays in the $P\overline{3}m1$ phase. However, the discussed dependencies of the band gap size on $x$ are somewhat different in Refs. \cite{Yu} and \cite{Hodgkins}. Furthermore, after the phase transition to the $P6_3/mmc$ type at higher $x$ values, the behavior of the band gap changes.

The spin-orbit coupling (SOC) is strong in Bi-based HaPs and significantly affects the electronic states in vicinity of the conduction band minimum (CBM) \cite{Pazoki,Ghosh}. Therefore, it is important to take SOC into account in the description of the electronic structure of HaPs. The density functional theory (DFT) calculations by Bass \textit{et al.} \cite{Bass}and Biswas \textit{et al.} \cite{Biswas} using hybrid functionals (HSE+SOC) obtained the indirect (direct) band gaps of Cs$_3$Bi$_2$Br$_{9}$ to be 2.52 (2.64) eV and 2.63 (2.75) eV, respectively, which are close to the experimental values \cite{Yu,Li,Jin}. In turn, the HSE+SOC calculations for Cs$_3$Bi$_2$I$_{9}$ yielded the indirect band gap of 2.3 eV \cite{Lehner,Rieger}. Unfortunately, no results of the HSE+SOC calculations in terms of the band gap sizes for the whole Cs$_3$Bi$_2$Br$_{9-x}$I$_x$ series were published so far, to our knowledge, only data from the no-SOC DFT calculations with the PBE (Perdew, Burke, and Ernzerhof \cite{Perdew}) functional are available \cite{Yu} for this series.

\begin{figure*}
\includegraphics[width=0.7\textwidth]{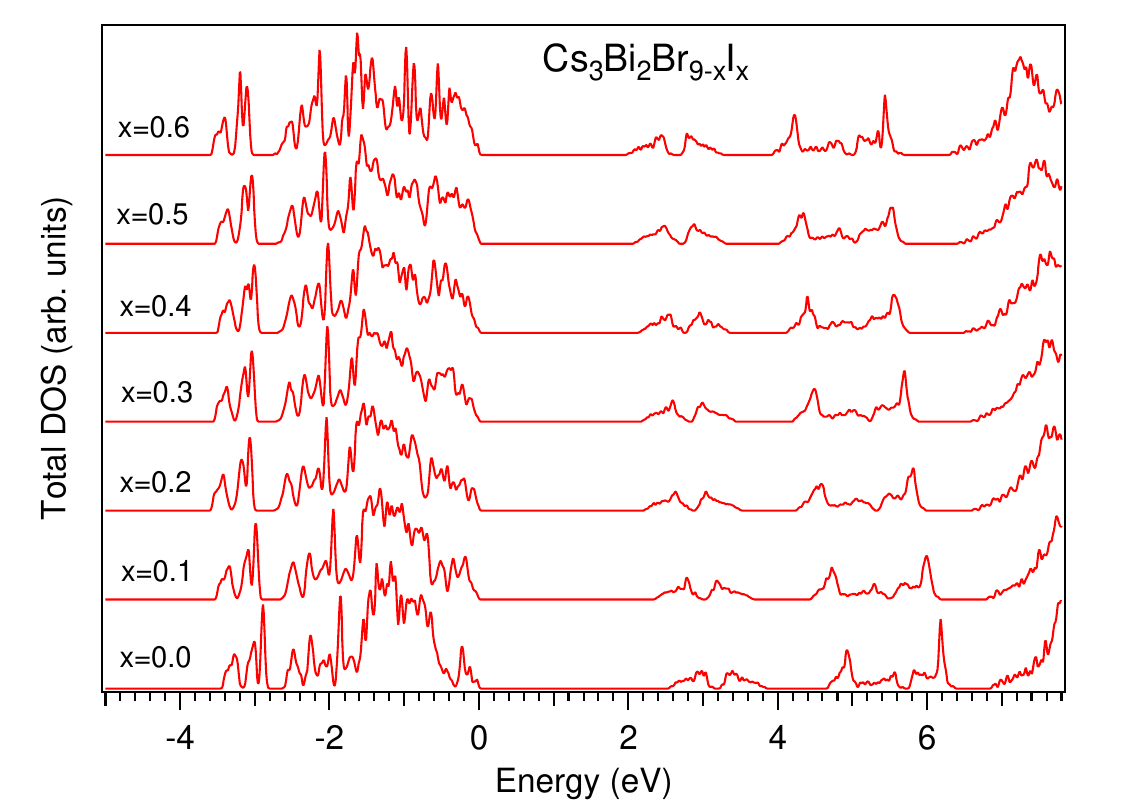}
\caption{Total density of states of Cs$_3$Bi$_2$Br$_{9-x}$I$_x$ calculated at the level of AK13/GAM theory. Zero eV is at the valence band maximum. \label{DOS_Cs3Bi2Br9-xIx}}
\end{figure*}

The intension of the present work was to investigate the electronic structure and the band gap variation in the Cs$_3$Bi$_2$Br$_{9-x}$I$_x$ series using the DFT method taking into account SOC. The recently suggested approach \cite{Butorin,Butorin2} was employed which combines the parameter-free Armiento-K\"{u}mmel generalized gradient approximation (AK13-GGA) exchange functional \cite{Armiento} with the nonseparable gradient approximation Minnesota correlation functional (GAM) \cite{Yu_GAM}. The approach allows for an efficient band gap estimation with accuracy similar to the HSE+SOC method but at the much lower computational cost when applying GGA functionals.

%Br9- HSE_SOC 2.63 (2.75) eV Biswas_2024 (indirect)
%Br9- HSE_SOC 2.52 (2.64) eV Bass_2016
%Br9- PBE 2.60 eV Wang_2021 (direct)
%Exp_Br9 2.57 (2.64) eV J.Jin-2023

%I9 HSE_SOC 2.3 eV Lehner_2015
%I9 HSE_SOC 2.32 eV Rieger_2019
%Exp_I9 2.1 (2.45) eV Ghosh_2017 (list)
%Exp_I9 1.85 (2.23) eV Yu_2019
%Exp_I9 1.98 (2.07) eV Gu_2018
%Exp_I9 1.9        eV Lehner_2015
%Exp_I9 2.2        eV Park_2015
%Exp_I9 2.8        eV Machulin_2004

%/////////////////////////////////////////////////////////
%-------------- Results and discussion -------------------
%/////////////////////////////////////////////////////////
\section{Results and discussion}
Fig.~\ref{DOS_Cs3Bi2Br9-xIx} displays the AK13/GAM-calculated total density of states (DOS) of Cs$_3$Bi$_2$Br$_{9-x}$I$_x$ for $x$=0.0, 0.1, 0.2, 0.3, 0.4, 0.5 and 0.6, i.e. when the system is in the $P\overline{3}m1$ phase. In each case, zero eV is at the valence band maximum (VBM). An inspection of the figure reveals that upon Br substitution by I the DOS shape stays the same in the conduction band and most of the valence band while some slight DOS changes are observed in the vicinity of VBM (within $\sim$1 eV region). The band gap decreases on going from $x$=0.0 to $x$=0.6. The width of the valence band slowly increases with $x$ due to extending DOS weight at VBM. Overall, the calculated DOS results of Cs$_3$Bi$_2$Br$_{9-x}$I$_x$ give an impression of the quasi-rigid-band behavior.

\begin{figure*}
\includegraphics[width=0.7\textwidth]{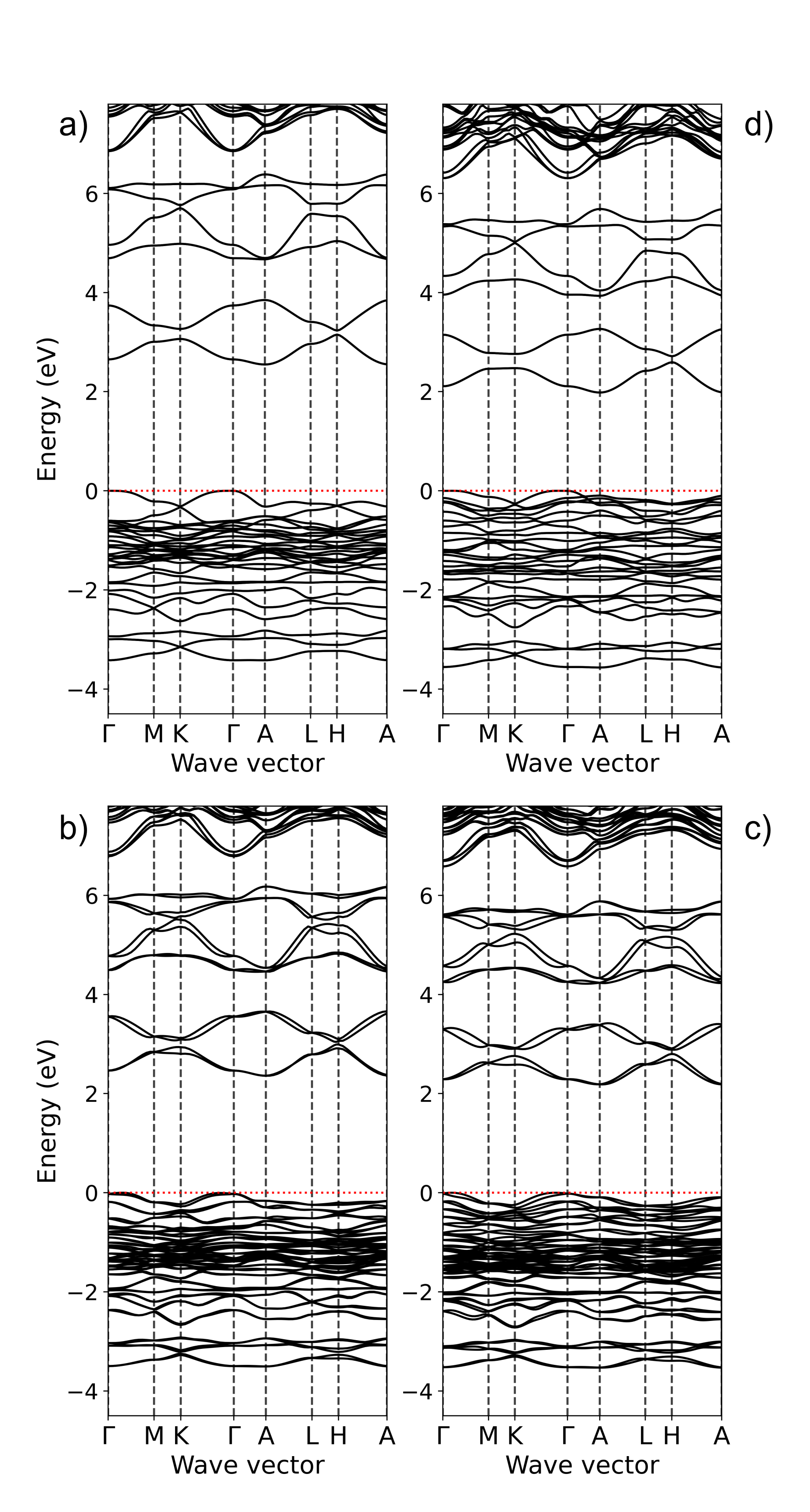}
\caption{Calculated band structure of Cs$_3$Bi$_2$Br$_{9-x}$I$_x$ for a) $x=0.0$, b) $x=0.1$, c) $x=0.3$, d) $x=0.6$. Zero eV is at the valence band maximum. \label{Bands}}
\end{figure*}

However, an inspection of the calculated band structure of Cs$_3$Bi$_2$Br$_{9-x}$I$_x$ in Fig.~\ref{Bands} suggests that the situation is somewhat more complex. As soon as Cs$_3$Bi$_2$Br$_{9}$ is doped with I, new bands appear very close to VBM. Further increase of the I content leads to a larger spread of the bands, thus "diluting" the compact band pack in the upper part of the valence band (compare Figs.~\ref{Bands}a and \ref{Bands}d).

\begin{figure}[t]
\includegraphics[width=\columnwidth]{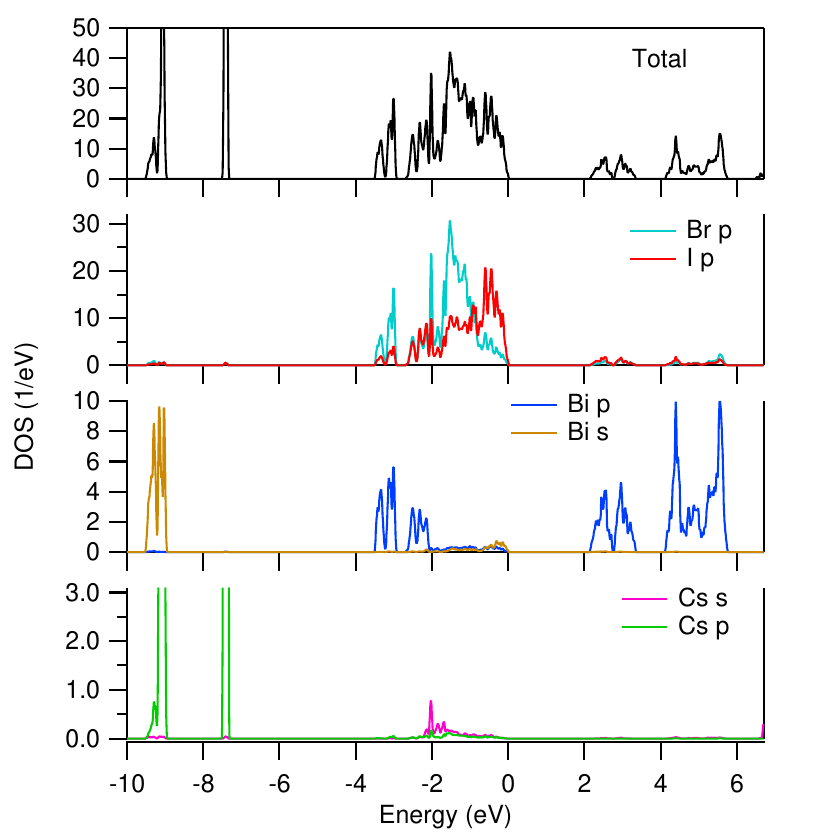}
\caption{Total and partial DOSs of Cs$_3$Bi$_2$Br$_{5}$I$_4$ calculated at the level of AK13/GAM theory. Zero eV is at the valence band maximum. \label{DOS_Cs3Bi2Br5I4}}
\end{figure}

The minimum band gap was calculated to be of indirect character for all the compositions. In the $P\overline{3}m1$-phase compounds, VBM and CBM are located at high symmetry points $\Gamma$ and $A$, respectively (see Fig.~\ref{Bands}). Due to a small difference between experimentally estimated indirect and direct band gaps \cite{Jin}, Cs$_3$Bi$_2$Br$_{9}$ was called as nearly direct band gap semiconductor. According to our calculations, such a difference in Cs$_3$Bi$_2$Br$_{9-x}$I$_x$ only slightly increases when going from $x$=0.0 to $x$=0.6. The indirect (direct) gaps are calculated to be 2.55 (2.65) eV and 2.02 (2.24) eV, respectively (Fig.~\ref{Bands}).

To characterize the contributions of all the elements to the valence and conduction bands, their calculated partial DOSs are shown in Fig.~\ref{DOS_Cs3Bi2Br5I4} for the Cs$_3$Bi$_2$Br$_{5}$I$_4$ composition. As one can see, the valence band is dominated by Br and I $p$ states. However, their distributions over the extent of the valence band are somewhat different. The gravity center of the Br $p$ DOS appears to be at lower energies and the I $p$ states provide the larger contribution in vicinity of VBM. As to Bi states in the valence band, the contribution of the Bi $s$ DOS at VBM seems to be not significantly larger than that of the Bi $p$ DOS which is in contrast to the situation in the Pb-based HaPs (where the Bi $s$ DOS dominates at VBM) also calculated at the level of AK13/GAM theory \cite{Butorin,Butorin2}. The conduction band region of Cs$_3$Bi$_2$Br$_{5}$I$_4$ between 2 and 6 eV in Fig.~\ref{DOS_Cs3Bi2Br5I4} is dominated by the contribution of the unoccupied Bi $p$ states. There are two groups of bands with their centroids at $\sim$2.8 and $\sim$5.0 eV, respectively. The splitting between them is a results of the spin-orbit coupling and hybridization with other states so that Bi $p$ DOS around 2.8 and 5.0 eV has mainly $6p_{1/2}$ and $6p_{3/2}$ character, respectively. In turn, the Cs $s$ and $p$ states are barely contributing to the valence band.

\begin{figure}[b]
\includegraphics[width=\columnwidth]{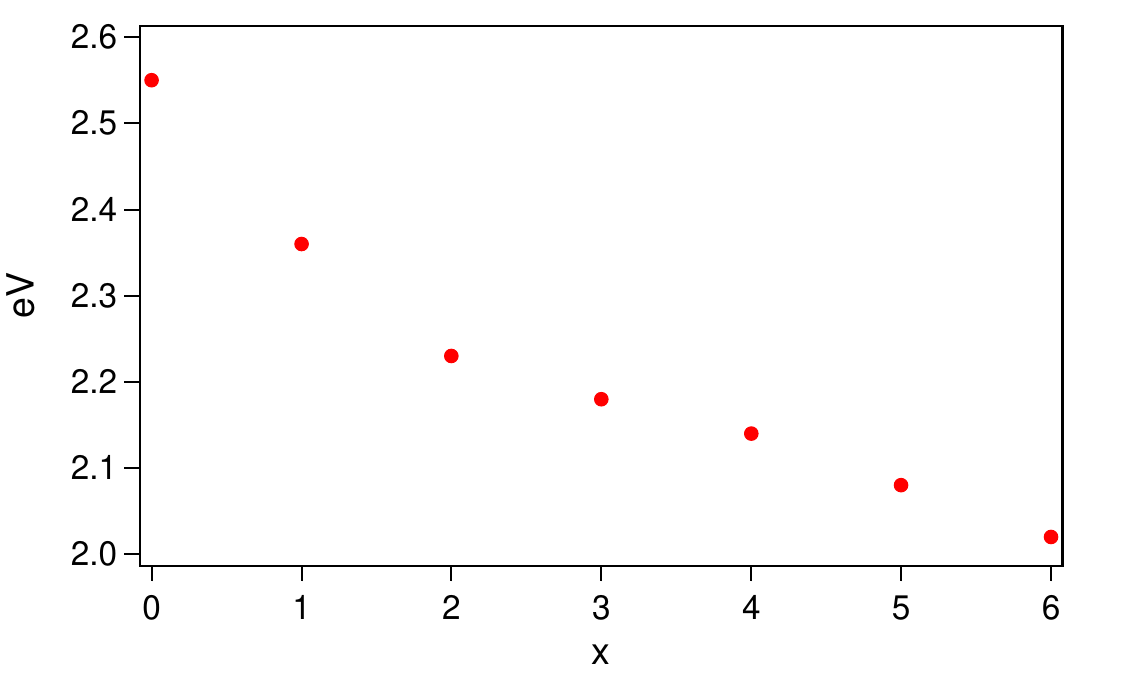}
\caption{Calculated band gap sizes of Cs$_3$Bi$_2$Br$_{9-x}$I$_x$ in the $P\overline{3}m1$ phase region. \label{Gaps}}
\end{figure}

Fig.~\ref{Gaps} displays the dependence of the AK13/GAM-calculated band gap size of Cs$_3$Bi$_2$Br$_{9-x}$I$_x$ on the $x$ value while the system is in the $P\overline{3}m1$ phase. The band gap decreases upon increasing I content but neither linear nor quadratic fits are very suitable to describe the calculated behavior. Nevertheless, this is in relatively good correspondence with available experimental data \cite{Hodgkins,Li}. At the same time, our results somewhat differ from those reported by Yu \textit{et al.} \cite{Yu} where the band gap sizes were calculated using the PBE functional without SOC and with the larger step for the $x$ value. The situation also differs from the case of the AK13/GAM calculations for MAPb(Br$_{1-x}$I$_x$)$_3$ (Ref.\cite{Butorin2}) where the band gap size dependence on the I content was described by the quadratic-like behavior.

\begin{figure}[h]
\includegraphics[width=0.8\columnwidth]{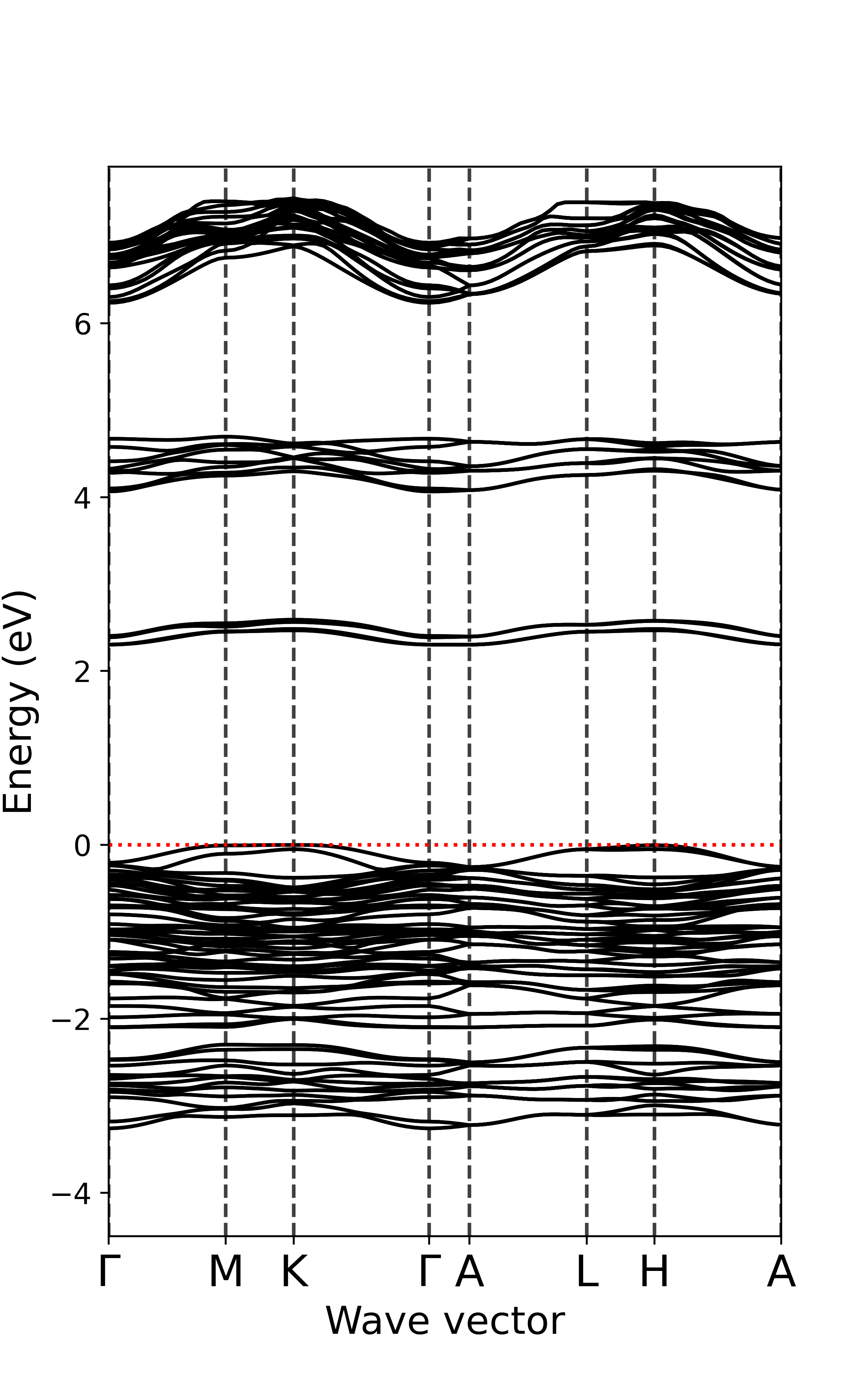}
\caption{Calculated band structure of Cs$_3$Bi$_2$I$_9$. Zero eV is at the valence band maximum. \label{Cs3Bi2I9_bands}}
\end{figure}

\begin{figure}[t]
\includegraphics[width=\columnwidth]{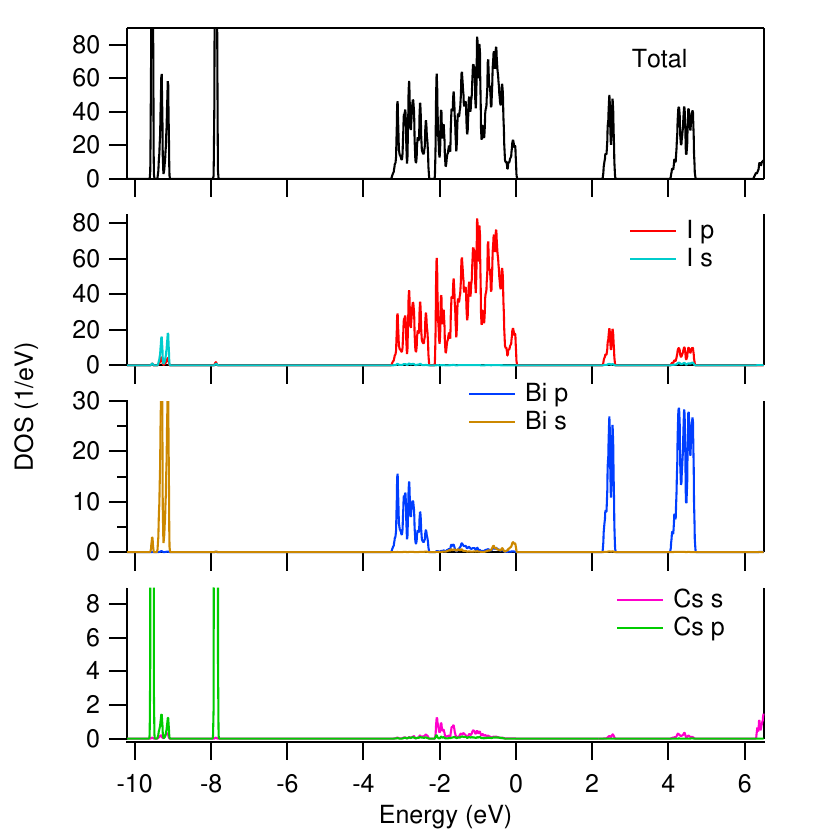}
\caption{Total and partial DOSs of Cs$_3$Bi$_2$I$_9$ calculated at the level of AK13/GAM theory. Zero eV is at the valence band maximum. \label{DOS_Cs3Bi2I9}}
\end{figure}

For Cs$_3$Bi$_2$I$_9$, both calculated DOSs and the band gap significantly change as compared to the Cs$_3$Bi$_2$Br$_{9-x}$I$_x$ series (up to $x$=0.6), which is not surprising, since Cs$_3$Bi$_2$I$_9$ has the $P6_3/mmc$ crystal structure. The AK13/GAM-calculated band structure of Cs$_3$Bi$_2$I$_9$ is shown in Fig.~\ref{Cs3Bi2I9_bands}. The band gap has indirect character and is located between high-symmetry points $K$ in the valence band and $A$ in the conduction band. The size of the band gap is found to be 2.30 eV which is an increased value and out of the trend for the decreasing band gap throughout the Cs$_3$Bi$_2$Br$_{9-x}$I$_x$ series from $x$=0.0 to 0.6 in the $P\overline{3}m1$ phase. It is difficult to compare this value with experimentally established band gap size of Cs$_3$Bi$_2$I$_9$ since the reported experimental data vary significantly as discussed above. However, our calculated value is in good agreement with other calculations using HSE+SOC \cite{Lehner,Rieger}. The different crystal structure of Cs$_3$Bi$_2$I$_9$ leads to a modification of its DOS (Fig.~\ref{DOS_Cs3Bi2I9}), particularly in the conduction band where the the Bi $6p_{1/2}$ and $6p_{3/2}$ sub-bands become significantly narrower.

As it was pointed out earlier \cite{Butorin}, the AK13/GAM approach allows for accurate calculations of shallow core-level energies which helps in the interpretation of the x-ray photoemission spectra (XPS). The binding energy of the Cs $5_{3/2}$ level in XPS of Cs$_3$Bi$_2$I$_9$ was reported to be $\sim$10.5 eV \cite{Phuyal}. The AK13/GAM-calculated position of this level is at 7.9 eV below VBM. If the size of the calculated band gap is added (7.9 + 2.3 = 10.2), the obtained value is very close to the measured binding energy for Cs $5_{3/2}$. This suggests that the chemical potential position is very close to CBM, thus indicating rather the $n$-type semiconductor character for Cs$_3$Bi$_2$I$_9$.

The inclusion of SOC is also important for the interpretation of the symmetry- and orbital-resolved x-ray spectroscopic data probing the unoccupied states of HaPs \cite{Phuyal,Man}.

%/////////////////////////////////////////////////////////
%--------------------- Conclusions -----------------------
%/////////////////////////////////////////////////////////
\section{Conclusions}

The application of the AK13/GAM method for the electronic structure calculations of Cs$_3$Bi$_2$Br$_{9-x}$I$_x$ in the $P\overline{3}m1$ phase provided the calculated band gap sizes similar to the experimental values. The band gap which is of the indirect character was shown to decrease with increasing I content and exhibited the non-linear dependence on the $x$ value. At the same time, the shape of the total DOS in the conduction band did not reveal any significant changes upon isovalent substitution in Cs$_3$Bi$_2$Br$_{9-x}$I$_x$.

%/////////////////////////////////////////////////////////
%------------------- Experimental ------------------------
%/////////////////////////////////////////////////////////

\section{Methods}

The DFT calculations were performed in the full-relativistic mode using the Quantum Espresso v.6.8 code \cite{Giannozzi}. A combination of AK13/GAM (as they are defined in the LibXC v.5.1.6 library \cite{Lehtola}) was applied for Cs$_3$Bi$_2$Br$_{9-x}$I$_x$ with varying $x$. The full-relativistic norm-conserving PBE pseudopotentials for cesium, bromine, iodine and bismuth were generated by the code of the ONCVPSP v.4.0.1 package \cite{Hamann} using input files from the SPMS database \cite{Shojaei}. The pseudopotentials were generated without non-linear core correction. An additional feature of this ONCVPSP version is its ability to check for positive ghost states. The valence configurations for the pseudopotentials were defined as 5s$^2$5p$^6$6s$^1$ for Cs, 4s$^2$4p$^5$ for Br, 5s$^2$5p$^5$ for I, and 5d$^{10}$6s$^2$6p$^3$ for Bi. The plane-wave cut-off energy was set to 60 Ry. The convergence threshold for density was 1.0x10$^{-12}$ Ry. The Brillouin zone was sampled using the Monkhorst-Pack scheme \cite{Monkhorst} and sizes of the shifted $k$-point mesh were chosen to be 8x8x6 for Cs$_3$Bi$_2$Br$_{9-x}$I$_x$ and 7x7x5 for Cs$_3$Bi$_2$I$_9$. The calculations were performed for the optimized crystal structures of Cs$_3$Bi$_2$Br$_{9-x}$I$_x$ adopted from Ref.\cite{Scanlon_collection} which were optimized using the PBEsol functional \cite{Perdew_PBEsol}. It has been shown \cite{Lindmaa} that AK13 is not accurate for this kind of the structural optimization. In the case of Cs$_3$Bi$_2$I$_9$, the crystal structure from Ref.\cite{Chabot} was used.

\textbf{Notes}
The author declare no competing financial interest.

%/////////////////////////////////////////////////////////
%------------------- Acknowledgements ---------------------
%/////////////////////////////////////////////////////////
\begin{acknowledgement}
The author acknowledges the support from the Swedish Research Council (research grant 2018-05525). The computations and data handling were enabled by resources provided by the Swedish National Infrastructure for Computing (SNIC) at National Supercomputer Centre at Link\"{o}ping University partially funded by the Swedish Research Council through grant agreement no. 2018-05973.
\end{acknowledgement}

%/////////////////////////////////////////////////////////
%--------------------- Bibliography ----------------------
%/////////////////////////////////////////////////////////
\bibliography{Cs3Bi2Br9-xIx}

\end{document}